\documentclass[twocolumn,longbib,twocolappendix]{aastex7}

\hypersetup{pdfauthor={Name}}
\usepackage{hyperref}
\usepackage{amsmath}

\begin{document}

\title{The Distribution of Earth-Impacting Interstellar Objects}
\shorttitle{Interstellar Impactors}
\shortauthors{Seligman, Mar{\v{c}}eta, \& Peña-Asensio}

\author[0000-0002-0726-6480]{Darryl Z. Seligman}
\altaffiliation{NSF Astronomy and Astrophysics Postdoctoral Fellow}
\affiliation{Dept. of Physics and Astronomy, Michigan State University, East Lansing, MI 48824, USA}
\email[show]{dzs@msu.edu}  

\author[0000-0003-4706-4602]{Du{\v{s}}an Mar{\v{c}}eta }
\affiliation{Department of Astronomy, Faculty of Mathematics, University of Belgrade, Studentski trg 16, Belgrade, 11000, Serbia}
\email{dusan.marceta@matf.bg.ac.rs}  
\author[0000-0002-7257-2150]{Eloy Peña-Asensio}
\affiliation{Department of Applied Mathematics and Aerospace Engineering, Universitat d'Alacant, 03690 Alacant, Spain}
\affiliation{Department of Aerospace Science and Technology, Politecnico di Milano, Via La Masa 34, 20156 Milano, Italy}

\email{eloy.pena@polimi.it}

\begin{abstract}

In this paper we calculate the expected orbital elements, radiants, and velocities  of Earth-impacting interstellar objects.   We generate a  synthetic population of $\sim10^{10}$ interstellar objects  with M-star kinematics   in order to obtain  $\sim10^4$ Earth-impactors. The relative flux of impactors arriving from the direction of the solar apex and the galactic plane is enhanced by a factor of $\sim2$  relative to the mean.  The fastest impactors  also  arrive from these directions, although Earth-impactors  are generally \textit{slower}  than  objects in the  overall  population. This is because the Earth-impacting subset contains a  higher fraction of low-eccentricity hyperbolic objects which are more strongly affected by gravitational focusing. Earth-impacting interstellar objects are more likely to have retrograde orbits close to the ecliptic plane. A selection effect makes the distribution of inclination of Earth-impacting interstellar objects  uniform/sinusoidal  at low/high perihelion distances. In turn, low perihelion impactors have higher impact probability towards the ecliptic plane. The overall impactor population therefore exhibits an intermediate inclination distribution between uniform and sinusoidal.  The highest velocity impacts  are most likely to occur in the spring when the Earth is moving towards the solar apex. However, impacts  in general  are  more likely to occur during the winter  when the Earth is located  in the direction of the antapex. Interstellar objects are more likely to impact the Earth at low latitudes close to the equator, with a slight preference for the Northern hemisphere due to  the  location of the  apex.   These distributions  are  independent of the assumed interstellar object number density, albedos, and  size-frequency distribution and are publicly available.
\end{abstract}


\section{Introduction} \label{sec:intro}

\added{The discovery of  1I/`Oumuamua \citep{Williams17}, 2I/Borisov \citep{borisov_2I_cbet}, and 3I/ATLAS \citep{Denneau2025}} demonstrated that large scale interstellar objects  traverse the Solar System somewhat routinely. The number density of interstellar objects was estimated based on nondetections prior to the discovery of these two objects \citep{sekanina1976probability, mcglynn1989nondetection,francis2005demographics,Moro2009,Engelhardt2014}.  The discovery of 1I/`Oumuamua implies a spatial number density of $\sim0.1$ au$^{-3}$  \citep{Jewitt2017,Trilling2017,Laughlin2017,Do2018,Levine2021} for $\sim100$ m  scale interstellar objects. However, the precise value is extremely uncertain and the size-frequency distribution of the population is entirely unconstrained.  

The \added{three} known interstellar objects exhibited divergent properties which are briefly summarized here. 1I/`Oumuamua was photometrically inactive \citep{Meech2017,Ye2017,Jewitt2017,Trilling2018}, although it had comet-like nongravitational acceleration \citep{Micheli2018}. While the high magnitude of nongravitational acceleration combined with lack of apparent comae was atypical, a population of  near-Earth objects with similar properties have  been reported \citep{Farnocchia2023,Seligman2023b,Seligman2024PNAS}.  1I/`Oumuamua had an extreme $6:6:1$ geometry inferred from its light curve \citep{Drahus2017,Knight2017,Fraser2017,McNeill2018,Belton2018,Mashchenko2019} and a moderately reddened  reflection spectrum \citep{Masiero2017,Fitzsimmons2017,Bannister2017}.  2I/Borisov  was discovered  in 2019 and displayed a dusty coma \citep{Jewitt2019b,Fitzsimmons:2019,Ye:2019,McKay2020,Guzik:2020,Hui2020,Kim2020,Cremonese2020,yang2021} with typical  carbon- and nitrogen-bearing species  \citep{Opitom:2019-borisov, Kareta:2019, lin2020,Bannister2020,Xing2020,Aravind2021} and an  enrichment of CO relative to H$_2$O \citep{Bodewits2020, Cordiner2020}. 1I/`Oumuamua and 2I/Borisov had  nuclear radii of $r_n\sim80$ m  and $r_n\sim400$ m respectively \citep{Jewitt2020}\footnote{While literature estimates of the nuclear radii of both objects vary, we report the average of all radii reported in Table 2 in \citet{Jewitt2023ARAA}.}.  \added{1I/`Oumuamua, 2I/Borisov, and 3I/ATLAS had  hyperbolic excess velocities of $V_\infty\sim26$ km s$^{-1}$, $V_\infty\sim32$ km s$^{-1}$, and $V_\infty\sim58$ km s$^{-1}$  corresponding to dynamical ages of $\sim10^2$, $\sim10^3$~Myr, and $\sim10$~Gyr \citep{Mamajek2017,Gaidos2017a, Feng2018,Fernandes2018,Hallatt2020,Hsieh2021,Taylor2025,Hopkins2025b}. 3I/ATLAS exhibited faint cometary activity and a reddened reflectance spectra similar to 1I/`Oumuamua ad 2I/Borisov \citep{Seligman2025,Opitom2025,Jewitt2025,Alarcon2025,Marcos2025,Kareta2025,Chandler2025,Frincke2025}. } We refer the reader to  \citet{Jewitt2023ARAA}, \citet{Fitzsimmons2023}, \citet{Seligman2023}, and \citet{MoroMartin2022} for  reviews on the  topic of interstellar objects

In theory, a subset of these interstellar objects traversing the inner Solar System should impact the Earth. However, only $\sim1-10$ total $\sim100$ m scale interstellar objects should have impacted the Earth since it formed \citep{Jewitt2023ARAA}. In general, interstellar impactors should be significantly less frequent than NEA impactors. Moreover,   distinguishing interstellar impact craters from solar system impact craters morphologically would be challenging \citep{Cabot2022}. These interstellar-originating craters may be easier to identify on the lunar surface and potentially  distinguishable based on their   small radii, low latitudes, and high melt fractions \citep{Chang2023}.

Smaller scale interstellar objects have been identified throughout the Solar System. For example, presolar grains in meteorites  provide  examples of small solids formed in extrasolar systems \citep{Zinner2014}. Interstellar dust particles have been detected in-situ by both the Ulysses and the Galileo spacecraft, implying a Solar System flux of $\sim10^{-3}-10^{-4}$ m$^{-2}$ s$^{-1}$  particles with $10^{-7}-10^{-6}$ m sizes \citep{Grun1993,Grun1997,Landgraf1998,Grun2000,Landgraf2000}.  Radar measurements obtained with the Arecibo Observatory  \citep{Mathews1999,Meisel2002a,Meisel2002b} constrained the flux of interstellar micrometeoroids (diameter $<10-100$ $\mu m$) to be $\sim10^{-8}$ m$^{-2}$ s$^{-1}$. The flux of $100$ $\mu m$ interstellar meteoroids was constrained by the  Canadian Meteor Orbit Radar (CMOR) \citep{Weryk2004} and $>20$ $\mu$m sized interstellar meteoroids by the Advanced Meteor Orbit Radar (AMOR) \citep{Baggaley1993,Baggaley2000,Taylor1996}.  We refer the reader to \citet{Sterken2019} for a recent review.

The search for larger scale interstellar impactors has been debated for over a century; however,  there remain no conclusive detections of interstellar meteors. In the early 1900s there was a claim that a substantial fraction of meteors in catalogues made by Von Niessl and Hoffmeister \citep{Niessl1925} were hyperbolic and therefore interstellar \citep{Fisher1928}. However, this was refuted in the subsequent decades \citep{Almond1951, Opik1956, Jacchia1961, stohl1970, Hajdukova1993, Hajdukova1994}.   Nondetections of interstellar meteoroids in optical data \citep{Hawkes1999} from the Canadian Automated Meteor Observatory (CAMO) \citep{Musci12} and the IUA Meteor Date Center \citep{Hajdukova2002,Hajdukova1994} have been used to constrain the overall flux. \citet{Wiegert2025} searched for interstellar events in the  Global Meteor Network (GMN) and found no compelling candidates.  While there are hyperbolic events in the Center for Near-Earth Object Studies databases \citep{PenaAsensio2022}, there appears to be no compelling evidence that these fireballs were interstellar \citep{Vaubaillon2022,BrownBorovicka2023,Hajdukova2024}.  It has been demonstrated that measurement errors --- and scattering via solar system planets in a small fraction of cases --- accounts for hyperbolic events in most optical meteor networks  \citep{,Wiegert2014,Hajdukova2014,Hajdukova2020a,Egal2017, Hajdukova2019, Hajdukova2020a, Hajdukova2020b} and of radar measured velocities of hyperbolic fireballs \citep{Baggaley07,Musci12}.

Hyperbolic fireballs have been recorded in other databases. One hyperbolic event was observed by the Finnish Fireball Network with 1-sigma confidence \citep{Eloy2024Icar40815844P} and fifteen were  observed by the European Fireball Network with 1-sigma confidence, including two with 3-sigma confidence \citep{Borovicka2022AA667A158B}. Five  candidate hyperbolic meteors have been identified with radar with 3-sigma confidence by CMOR \citep{Froncisz2020PSS19004980F}. All these cases, however, are  close to the parabolic limit, unlike 1I/`Oumuamua and 2I/Borisov. Hyperbolic meteors are also reported in modern automated optical meteor network: hyperbolics make up i) 8\% of all meteors recorded by the Global Meteor Network (GMN) \citep{Vida2021MNRAS5065046V} (updated on Jan 2025) and ii)  12\% of all meteors in the Meteoroid Orbit Database v3.0 by Cameras for All-Sky Meteor Surveillance (CAMS) \citep{Jenniskens2018PSS15421J}. However, the researchers leading these networks have never claimed the detection of interstellar impactors. 

The incoming extrasolar population remains poorly constrained across a broad size range, as linking interstellar dust and kilometer-scale objects to estimate the expected number of hyperbolic meteoroid impacts is not supported by meteor surveys  \citep{PenaAsensioSeligman2025}. However, there have been efforts to calculate the distributions of interstellar objects passing through the Solar System. \citet{Cook2016} and \citet{Engelhardt2017} produced some of the first synthetic populations of interstellar objects via direct numerical  integrations. Other efforts to generate Monte Carlo simulations of interstellar objects include \citet{Seligman2018,Hoover2022,Stern2024}. \citet{Marceta2023a} introduced the ``probabilistic method'' to generate synthetic interstellar objects. This method is   orders of magnitude more computationally efficient than alternative approaches because it propagates trajectories analytically.   It has been implemented to make predictions regarding the distribution of interstellar objects
that should be detected with the forthcoming Rubin Observatory Legacy Survey of Space and Time (LSST) \citep{Marceta2023b,Dorsey2025} for a variety of assumptions regarding their kinematics \citep{Hopkins2023,Forbes2024,Hopkins2025}.

In this paper we implement the probabilistic method to generate a population of Earth-impacting interstellar objects.  The orders of magnitude increase in computational efficiency of the probabilistic method   enables  the generation a statistically robust synthetic population of Earth-impacting interstellar objects. Specifically, we generate a population of $>10^{10}$ interstellar objects that give us a population of $>10^5$ Earth-impactors.  We do not intend to make any claims about the presence or lack of interstellar meteors in extant data. Instead, these distributions are intended to be a useful and open-source tool to investigate Earth-impacting interstellar objects of macroscopic sizes, i.e., governed solely by gravity and unaffected by radiation.

 \begin{figure*}
\includegraphics[width=1.\linewidth]{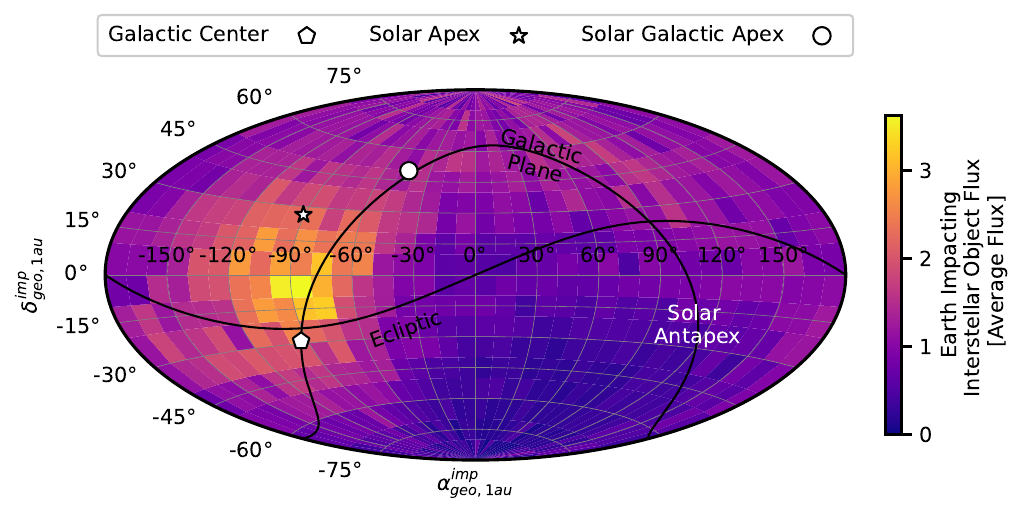}
\caption{The  radiants towards the Earth of impacting interstellar objects.  Interstellar objects tend to impact the Earth in the directions of the solar apex and the galactic plane.  Radiants are shown in the geocentric frame.}
\label{fig:skymap}
\end{figure*}

\section{Synthetic Population}\label{sec:simulations}

In this section we describe the methodology to generate the synthetic population of interstellar objects and impactors. We implement the probabilistic method that was developed by \citet{Marceta2023a} to generate all synthetic objects. The advent of the probabilistic method   enables  generation of a sufficient number of synthetic interstellar objects to obtain robust statistics on the subset of Earth-impactors.

\subsection{Assumptions}
In these calculations we purposefully do \textit{not} assume a (i) size-frequency distribution, (ii) albedo distribution or, (iii) spatial number density of interstellar objects --- all of which are entirely unconstrained.  In other words, we do not attempt to predict the rate of interstellar impactors; we only calculate their expected and \textit{unnormalized} distribution. It is worth noting that there are hints at what the size-frequency distribution of interstellar objects may be. For example \citet{LecavelierdesEtangs2022} inferred a similar size-frequency distribution of exocomets in the $\beta$ Pictoris system as in Solar System populations of small bodies \citep{Boe2019,Tancredi2006,Snodgrass2011,Meech2004,Fernandez2013,Bauer2017}. However, it is unclear if the size-frequency-distribution of bound exocomets would be representative of the ejected subset. That being said, the distributions presented here are only applicable to interstellar objects that are large enough to not be affected by gas drag in the interstellar medium  \citep{Draine2011,MoroMartin2022_seeds}.

We assume that the interstellar objects have the same galactic kinematic    as   M-stars. This choice is admittedly somewhat arbitrary because the  kinematics of interstellar objects  is  unconstrained. A detailed comparison of impactor properties for various galactic kinematics is outside of the scope of this work.

\subsection{Orbit Generation}

We calculate the subset of Earth-impacting interstellar objects as follows.  \added{First, test populations were generated within 1 au and propagated through time under several assumed kinematic scenarios to determine the timescale for objects to leave this region. The residence time  was found to be about 80 days in all cases. This duration represents the maximum time during which any ISO within 1 au could impact Earth. Next, the maximum heliocentric distance from which an ISO could reach 1 au within 80 days was calculated, assuming a speed of 100 km/s directed toward the Sun, yielding 4.5 au. Consequently, the full synthetic population was generated within 5.5 au. Finally, for each ISO orbit, the Minimum Orbit Intersection Distance (MOID) with Earth’s orbit was computed, and only those objects satisfying two conditions were retained: (1) their MOID was smaller than Earth’s effective radius, and (2) their orbital position allowed them to reach the intersection point within 80 days, corresponding to the approximate dynamical stationarity of the population near 1 au. In other words, interstellar objects in our synthetic population are considered impactors if they intersect the torus centered on the Sun with major/minor radius of $1$au/$1R_\oplus^{\rm eff}$ respectively,  where $R_\oplus^{\rm eff}$ is the effective radius of the Earth, accounting for Earth's gravitational focusing, and is defined as
\begin{equation}
R_\oplus^{\rm eff} = R_\oplus \, \sqrt{1 + \left(\frac{v_{\rm geo}^{\rm esc}}{v_{\rm geo}^{\rm imp}}\right)^2},
\end{equation}
where $R_\oplus$ is the physical radius of the Earth, $v_{\rm geo}^{\rm esc}$ is Earth's escape velocity, and $v_{\rm geo}^{\rm imp}$ is the impact velocity. The effective radius is calculated individually for each impact.
 For the remainder of this paper we refer to this volume as the ``Earth-torus'' for lack of a better nomenclature. 
  \begin{figure*}
\includegraphics[width=1.\linewidth]{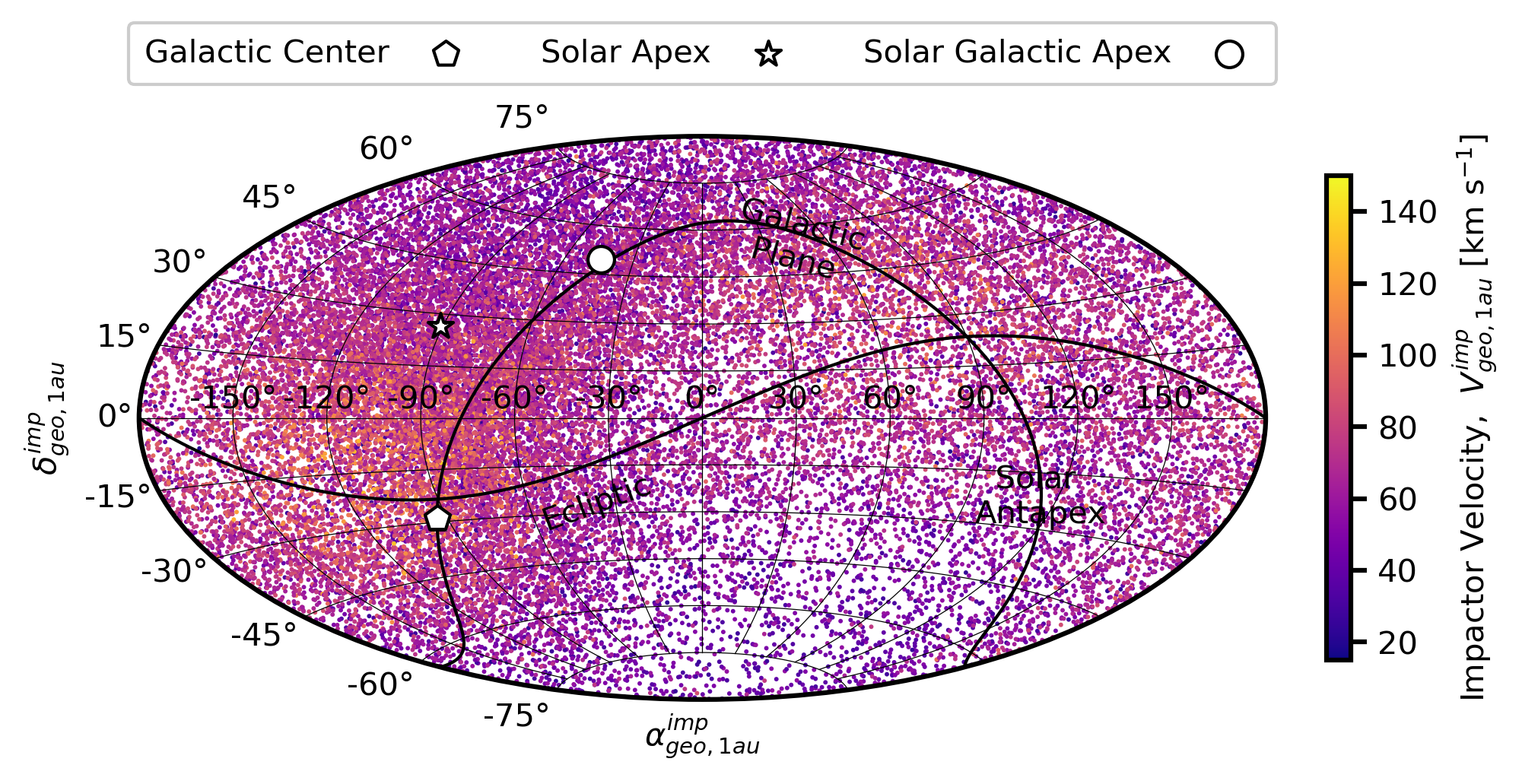}
\caption{Interstellar objects impact the Earth with higher velocity when approaching from the solar apex and the galactic plane.  Impactor velocity is calculated in  the geocentric frame. }
\label{fig:skymap_velocity}
\end{figure*}

Therefore, our method generates $\sim1$ impactor for every $\sim8\times10^5$ synthetic objects. Robust statistics at the population level were achieved with a sample of Earth-impactors on the order of $\sim3.3\times10^4$, enabling the identification of relevant structures in the population. This required the computation of $\sim2.6\times10^{10}$ synthetic interstellar objects, which demonstrates the prerequisite for the computational efficiency of the probabilistic method.}

The probabilistic method  incorporates  gravitational focusing from the Sun, but \textit{not} from the Earth. Technically, the mutual gravity from both the Earth and the Sun  should be incorporated via direct numerical integration to evaluate this population. However,  this is not  feasible  computationally for  the  total number of synthetic objects required to generate a substantive impactor population. However, meteor databases typically provide both the geocentric and apparent/observed radiants and velocity, as well as osculating orbital elements. Therefore, the distributions presented in this paper will still be straightforward to compare with meteor databases.

\subsection{Terminology}
We briefly describe some terminology here that we will implement throughout the remainder of this paper to avoid confusion. We differentiate between the geocentric and heliocentric frame for a given parameter $X$ with subscripts $X_{geo}$ and $X_{hel}$. In this paper, heliocentric means ecliptic Sun-centered while geocentric means equatorial Earth-centered. We distinguish between a parameter that has been affected (at 1 au) and has not been affected by gravitational focusing of the Sun with subscripts $X_{,1au}$ and $X_{,\infty}$ respectively. Finally, we differentiate between the interstellar impactors and the overall interstellar population with the superscript $X^{imp}$ and $X^{all}$. For example, the velocity of the impactor population in the geocentric frame at infinity would be $V_{{geo},\infty}^{imp}$.

All of the eccentricities described throughout the paper are hyperbolic by construction. We routinely describe eccentricities that are close to 1 ($e\sim1-2$) as low-eccentricity, not to be confused with near-circular ($e\sim0$).

Finally, in this paper we describe velocity vectors and magnitudes often and interchangeably. For the remainder of this paper, a quantity is only assumed to be a vector if it has the vector arrow denotation. For example $\vec{V}$ is the velocity vector, but $V$ denotes only the magnitude of the vector. 

\begin{figure*}
\includegraphics[width=1.\linewidth]{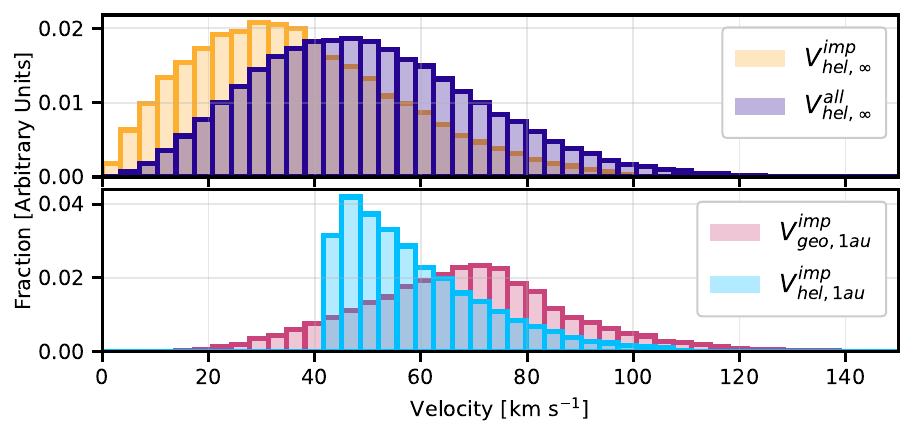}
\caption{The distribution of velocities of Earth-impacting interstellar objects are distinct from the overall interstellar object population. The addition of the Earth's velocity shifts the lognormal distribution to higher median values for the geocentric frame.}
\label{fig:velocity}
\end{figure*}

\section{Distribution of Interstellar Impactors}\label{sec:distribution}
In this section we present the distribution of  Earth-impacting interstellar objects. In the following three subsections we describe the orbits, seasonal variability, and Earth-locations of the population.

\subsection{Orbits of Impactors}
 In Figure \ref{fig:skymap} we show  the relative flux of Earth-impacting interstellar objects  as a function of their radiants in geocentric equatorial J2000. There are flux enhancements/deficits of a factor of $\sim2$ compared to the mean in the direction of the solar apex/antapex. There is also an enhancement of impactors in the direction of the galactic plane.  The faster objects tend to cluster in the  direction of the solar apex and the galactic plane (Figure \ref{fig:skymap_velocity}). 
 
 The cells shown in Figure \ref{fig:skymap} are evenly spaced in right ascension ($\alpha$) and declination ($\delta$), and  therefore have  \textit{uneven} surface area. We correct for this geometric effect by  dividing  the  number of objects in each cell by its normalized surface area, $\sigma_{ij}$, given by:
\begin{equation}\label{eq:area}
    \sigma_{ij} = |\alpha_{i+1/2}-\alpha_{i-1/2}|  |\sin{\delta_{j+1/2}}-\sin{\delta_{j-1/2}}|
\end{equation}
In Equation \ref{eq:area},    the subscripts $i$ and $j$ correspond to the horizontal and vertical indices of the grid respectively. \added{Figures \ref{fig:skymap} and \ref{fig:skymap_velocity} use an Aitoff projection.}

In Figure \ref{fig:velocity} we show the distributions of velocities of the Earth-impactors and of the entire interstellar population. The  impactors   peak at geocentric velocities of $\sim 72$ km s$^{-1}$\added{. This corresponds to  the maximum collisional velocity of two solar system objects which interstellar particles may exceed \citep{Hajdukova2020b}.} The interstellar population should encounter the Solar System with the mean velocity of the Sun with respect to the local standard of rest. Therefore, the heliocentric velocity distribution at 1 au should peak at  $\sim45$ km s$^{-1}$ after considering the solar acceleration.  The slowest impactors should encounter the Earth from behind with geocentric velocities of $12$ km s$^{-1}$ ($42-30$ km s$^{-1}$)\footnote{The  escape velocity of the Solar System at 1 au minus the heliocentric velocity of the Earth.}. 

It is  somewhat surprising that the geocentric impactor distribution  peaks  at $\sim72$ km s$^{-1}$. Intuitively, we would expect that the geocentric velocities of $\sim75$ ($45+30$ km s$^{-1}$) and $\sim15$ km s$^{-1}$ ($45-30$ km s$^{-1}$) would have roughly the same probability. However, it is much more likely for an interstellar impact to have higher velocities. The relative collision velocity, $ \vec{V}_{rel,1au}$ (which is effectively our $V_{{geo},1au}^{imp}$), is given by:

\begin{equation}\label{eq:vrel}
|\vec{V}_{rel,1au}|=\sqrt{|\vec{V}_{hel,1au}|^2-2|\vec{V}_{hel,1au}|| \vec{V}_{\oplus}|\,\cos{\psi}\,+|\vec{V}_{\oplus}|^2}\,.
\end{equation}
In Equation \ref{eq:vrel}, $\vec{V}_{hel,1au}$ and $\vec{V}_{\oplus}$ are the heliocentric velocity vectors of the interstellar object and the Earth respectively, while $\psi$ is the angle between them. It is evident from Equation \ref{eq:vrel} that  a collision with low relative velocity  only occurs when the Earth and the interstellar object move in the same direction, or for small $\psi$. In other words, low relative velocities only occur when the interstellar object moves in the ecliptic plane and directly impacts the Earth from behind. As the relative velocity increases, the number of possible combinations that can result in a given relative velocity increases. Therefore the distribution of velocities is naturally skewed towards higher values.

\begin{figure}
\includegraphics[width=1.\linewidth]{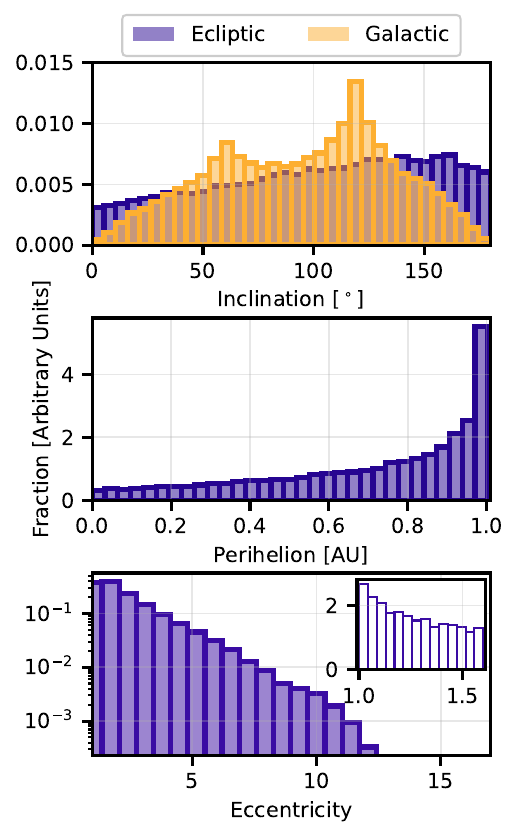}
\caption{Earth-impacting interstellar objects are significantly more likely to have perihelia close to the Earth  and  hyperbolic orbits near the low-eccentricity limit. The inclination distribution measured with respect to the ecliptic is uniform in angular space. In physical space, there is a larger fraction of impactor orbits in the ecliptic plane. This is because prograde and retrograde ecliptic interstellar objects are significantly more likely to cross the Earth-torus. The inset in the lower panel shows the low-eccentricity limit  and has the same axes.}
\label{fig:perihelia}
\end{figure}

 The median \textit{heliocentric} velocity of the impactors is slower than that of the overall interstellar population. We reiterate for clarity  that in this paper heliocentric means ecliptic Sun-centered while geocentric means equatorial Earth-centered. This excess of slowly moving impactors can be readily explained. These objects have low-eccentricity  orbits; it is evident in Figure \ref{fig:perihelia} that the vast majority of objects have eccentricity close to $e=1$. These low-eccentricity orbits have corresponding slow velocities at infinity, $V_{hel,\infty}\simeq0$ km s$^{-1}$ (Figure \ref{fig:velocity_eccentricity}). These  objects spend more time in a given heliocentric range than faster objects and therefore experience stronger  gravitational focusing. For example,  an object with $q=1$ au and $V_{hel,\infty}=0$ km s$^{-1}$ spends  $\sim30\%$ more time in the vicinity of the Earth compared to an object with q=1 au and $V_{hel,\infty}=25$ km s$^{-1}$. Therefore, low-eccentricity objects are more likely to enter the Earth-torus.

In conclusion, the slow impactors  have essentially the same velocity at 1 au ($\sim$ 42-43 km s$^{-1}$) even if they have differences of up to 9 km s$^{-1}$ at infinity. This implies that the majority of interstellar objects  that impact the Earth should do so just slightly above the escape velocity from the Solar System at 1 au. It is worth noting that our choice of adopting M-star kinematics diminishes this effect; if we had used the O/B or G kinematics, the distribution would be even more concentrated towards the low-eccentricity limit because they are slower.

 \begin{figure}
\includegraphics[width=1.\linewidth]{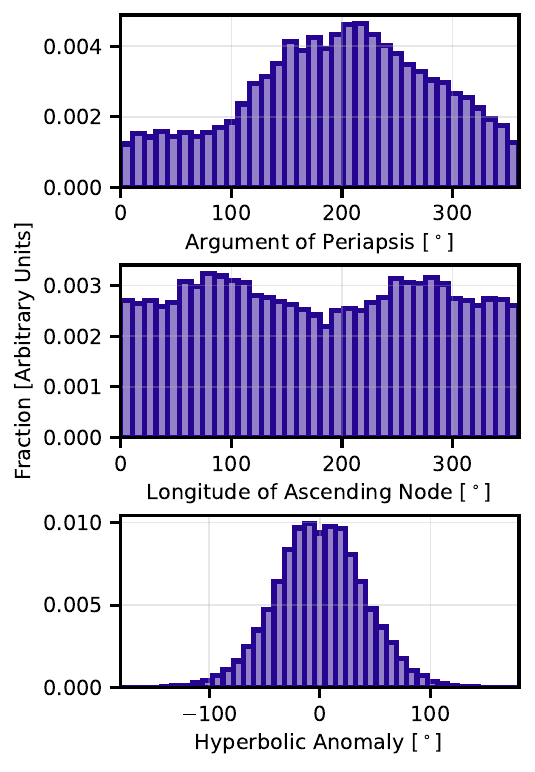}
\caption{The distributions of longitude of ascending node, argument of periapsis, and hyperbolic anomaly of the impactors are approximately symmetric. These distributions are similar to those in the overall interstellar populations \citep{Marceta2023a}. }
\label{fig:angles}
\end{figure}
 \begin{figure}
\includegraphics[width=1.\linewidth]{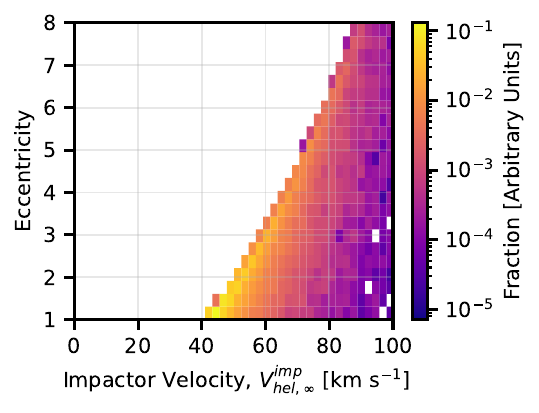}
\caption{The distribution of velocity of impactors correlates with eccentricity. The low-eccentricity objects all have low velocities at infinity. Note that the zoomed-in axes do not show the entire range of eccentricities and velocities in the impactor population. }
\label{fig:velocity_eccentricity}
\end{figure}

In Figures \ref{fig:perihelia} and \ref{fig:angles} we show the orbital elements of the Earth-impactors.   The majority of Earth-impactors have low-eccentricity orbits with  perihelia close to the Earth-distance from the Sun. The prevalence of $q\sim1$ au perihelion distances was explained in \citet{Marceta2023a}. The $q\sim1$ au impactors in our population correspond to the outer limiting trajectory in Figure 1 of that paper, where the radial velocity in the shell is zero. The relative number density is obtained in Equation 12 of that paper by dividing the flux by the area of the shell and the radial velocity. This results in a  predominance of orbits with perihelion close to the  outer radius of a given shell for  every shell.  These outer most  orbits are therefore the slowest moving objects in the radial direction (for a given shell), meaning they spend the most time inside the shell compared to orbits with lower perihelia. Therefore, they are the most likely orbits to impact.

\begin{figure}
\includegraphics[width=1.\linewidth]{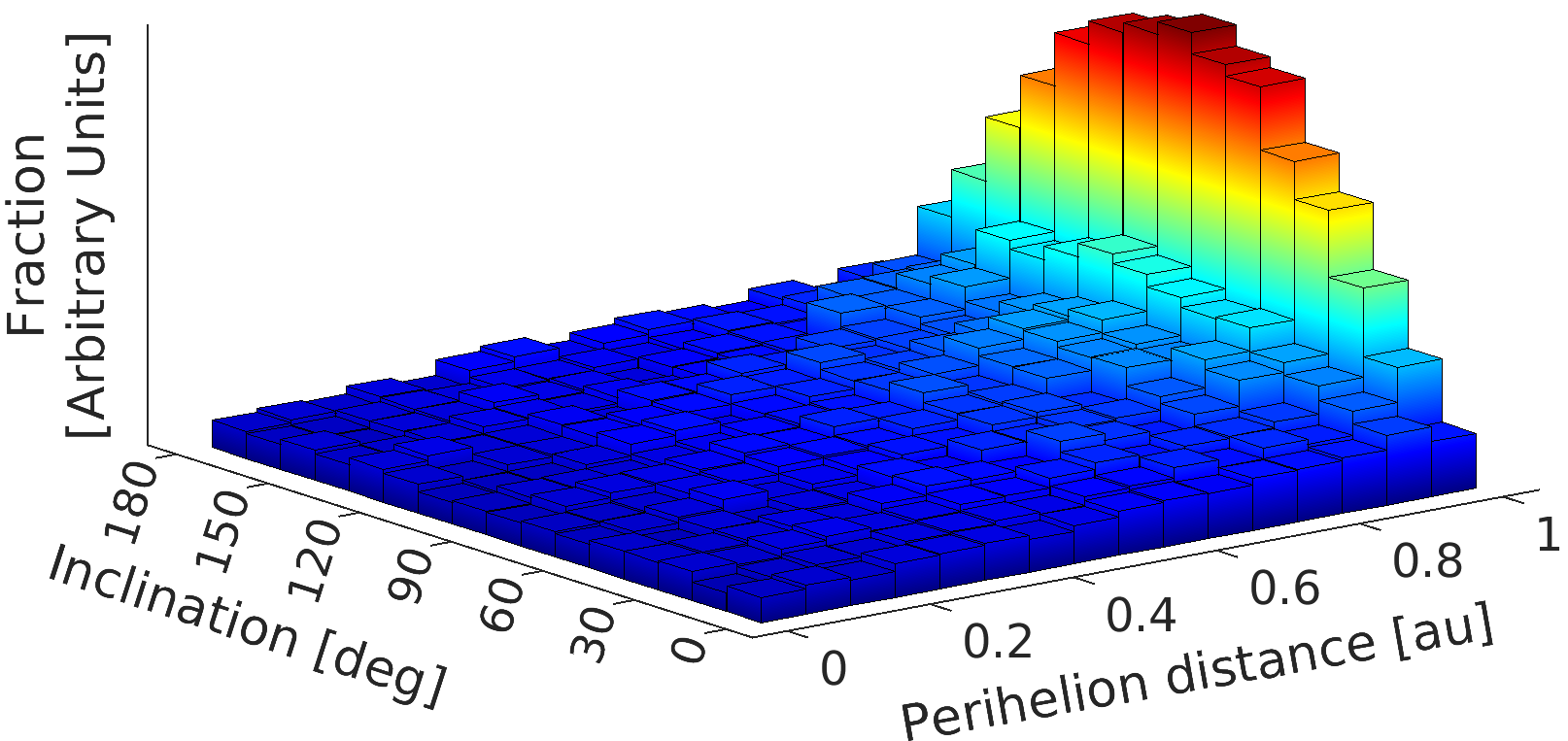}
\caption{\added{Unweighted inclination distribution of earth-impacting interstellar objects becomes sinusoidal/uniform at larger/smaller perihelion distances.}}
\label{fig:hist3d}
\end{figure}

\added{The inclination distribution of impactors differs significantly from that of the overall population. As shown in Figure~\ref{fig:hist3d}, it is closely linked to the perihelion distance. This bivariate histogram shows only the distribution of all objects that entered the Earth-torus during the analyzed period, without accounting for their impact probability, which is discussed later in this section. With increasing perihelion distance, the inclination distribution approaches a sinusoidal form, consistent with the overall population. Conversely, for smaller perihelion distances, the distribution becomes increasingly uniform.}

\begin{figure}
\includegraphics[width=1.\linewidth]{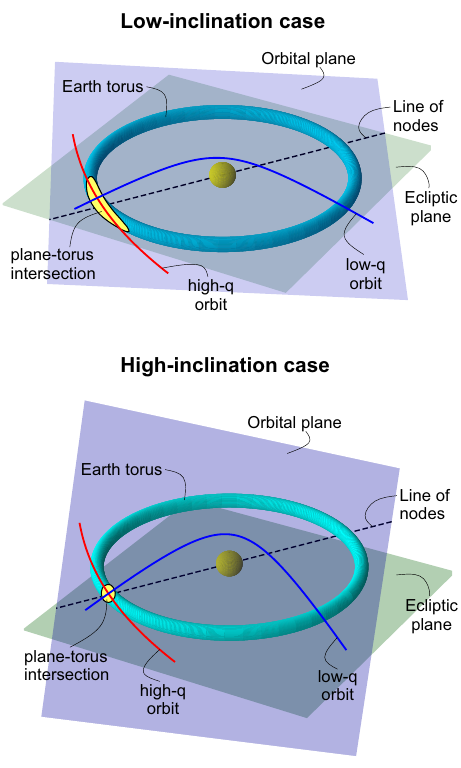}
\caption{Geometrical selection effect illustrated for low inclination (upper panel) and high inclination (lower panel). Low-$q$ orbits intersect Earth’s path nearly perpendicular, while high-$q$ orbits intersect nearly tangentially. The length of the intersection arc depends only on inclination, having a strong effect for low-$q$ orbits but a negligible effect for high-$q$ orbits.}
\label{fig:selection_effect}
\end{figure}

\added{This behavior arises from a geometrical selection effect illustrated in Figure ~\ref{fig:selection_effect}. For small values of $q$, the orbit intersects Earth’s orbital path almost perpendicular. In this regime, the range of argument of perihelion values that allow an intersection depends strongly on inclination. At low inclinations, the intersection arc along Earth’s orbit is longer, permitting a wider range of argument-of-perihelion values that produce a crossing.
For larger $q$, the orbit intersects Earth’s orbital path more tangentially, and the impact condition becomes less sensitive to inclination, since similar range of argument of perihelion values still satisfies the intersection criterion. As a result, low-inclination orbits are favored, and the resulting inclination distribution becomes flatter than the original one.}

\added{An additional reason why orbits close to the ecliptic are favored is that objects on such orbits have a longer residence time within the Earth-torus. This increases their likelihood of impact, as they can encounter Earth over a larger fraction of its orbit. Furthermore, since the impact flux is proportional to $v_{\mathrm{geo}, 1,\mathrm{au}}^{\mathrm{imp}}$, all distributions are skewed toward retrograde objects, as is clearly visible in the top panel of Figure \ref{fig:perihelia}. To account for this effect, we propagated the orbits of all impactors and calculated their residence time ($t_{\mathrm{res}}$) inside the Earth-torus. Then, all the distributions shown in Figures \ref{fig:perihelia} and \ref{fig:angles} were weighted by $t_{\mathrm{res}} \cdot v_{\mathrm{geo}, 1,\mathrm{au}}^{\mathrm{imp}}$.}

The inclination distribution of impactors  with respect to the galactic plane  has  \added{two peaks} at   $i\sim60^\circ$ and $i\sim120^\circ$ \added{(Figure \ref{fig:perihelia})}. These two peaks  correspond to direct and retrograde interstellar objects in the ecliptic plane whose obliquity to the galactic plane is $60^\circ$. \added{The right peak is more pronounced because it corresponds to retrograde objects. This occurs because the impact flux is proportional to $v_{geo}$, as mentioned. This effect also causes the inclination distribution with respect to the ecliptic plane to be asymmetric toward retrograde objects.}

The remaining three orbital element angles (Figure \ref{fig:angles}) are roughly symmetric. The distributions of argument of periapsis and longitude of the ascending node are bimodal. All three distributions are  similar to those  for the overall interstellar population, as can be seen upon visual comparison with Figure 9 in \citet{Marceta2023a}.

\subsection{Seasonal Dependence}

\begin{figure}
\includegraphics[width=1.\linewidth]{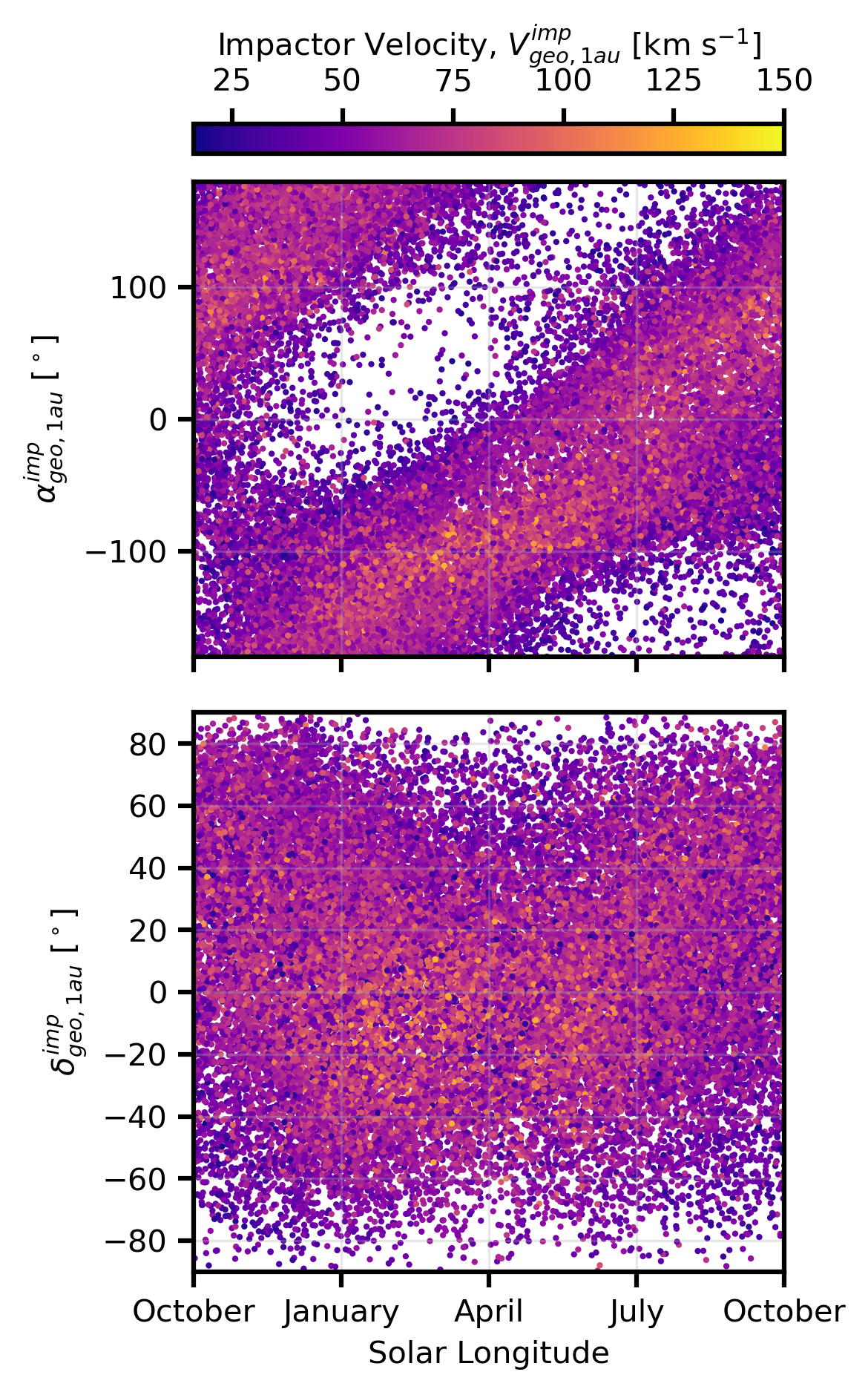}
\caption{Faster interstellar objects are more likely to impact the Earth in the spring when the Earth is moving towards the apex. The declination distribution of high velocity impactors mirrors the ecliptic plane. Velocities shown are calculated in the geocentric frame}
\label{fig:solarlong}
\end{figure}

In this subsection we present the seasonal dependance of the interstellar impactor distributions. In Figure \ref{fig:solarlong} we show the radiants of   Earth-impacting interstellar objects as a function of solar longitude and velocity. The calendar month is calculated assuming that the Earth has the same ecliptic longitude as the interstellar object when it crosses the Earth-torus. The seasons are defined based on the Northern hemisphere.   To convert the solar longitude to month, we neglect the fact that the solar longitude is slowly drifting with respect to the seasons. For simplicity, we plot October, January, April, and July as $\pm\pi,-\pi/2,0$, and $\pi/2$.

There is significant substructure in the parameter space spanning radiants, impactor velocity,  and solar longitude implying significant seasonal dependence of impactors. For example, the fastest interstellar objects impact  in the spring when the Earth is moving towards the apex.

\begin{figure}
\includegraphics[width=1.\linewidth]{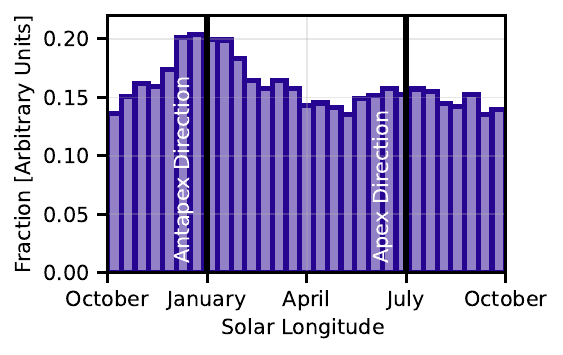}
\caption{Interstellar objects are more likely to impact the Earth in the winter than the spring.}
\label{fig:solarlong_histogram}
\end{figure}

However, interstellar objects are overall more likely to impact when the Earth is  in the direction of the antapex which occurs in winter. In Figure \ref{fig:solarlong_histogram} we show the distribution of total relative impacts as a function of solar longitude. The flux increases during winter, when the Earth is near the solar antapex, as the Sun acts as a gravitational lens that focuses a larger number of particle trajectories toward Earth’s orbit, in agreement with \citet{Strub2019AA621A54S}.

\begin{figure}
\includegraphics[width=1.\linewidth]{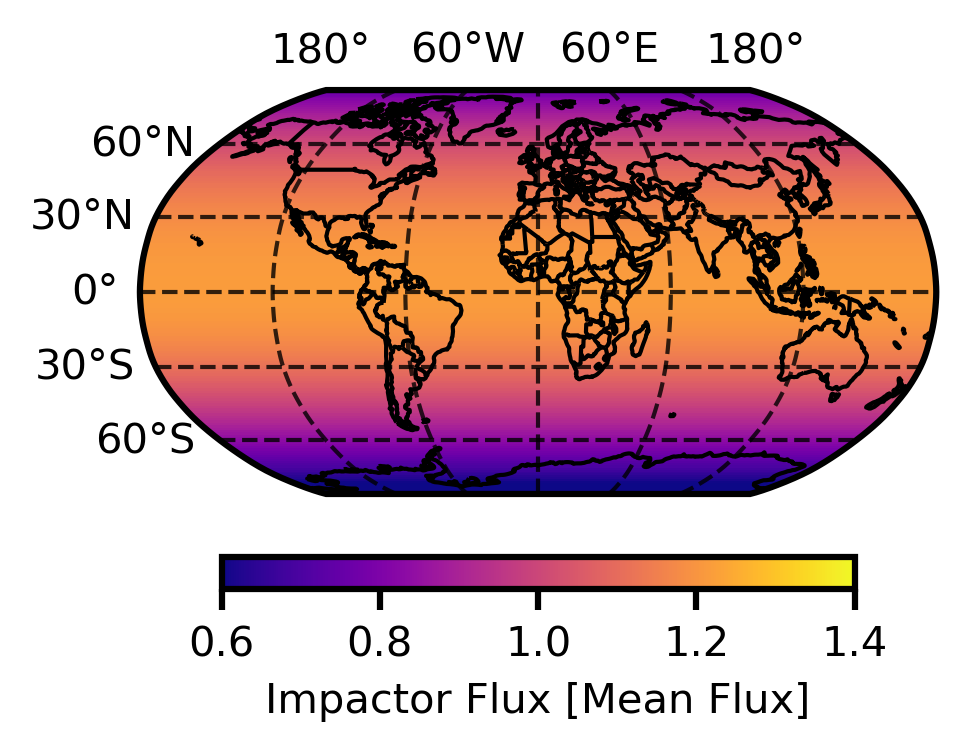}
\caption{Interstellar objects  are more likely to impact the Earth at low latitudes close to the equator. There is a slight preference for impactors in the Northern hemisphere. }
\label{fig:worldmap_flux}
\end{figure}

We provide a discussion of the physical explanation for this bias. Consider two trajectories of interstellar objects arriving in the direction of the apex on the near and far side of the Sun compared to the  motion of the Earth. An interstellar object coming from the far side of the Sun in the apex direction has a higher chance of impact because it collides after perihelion. This means it can originate from a slightly larger impact parameter; moreover,  the number of interstellar objects is linearly proportional to the impact parameter \citep{Marceta2023a,Seligman2023}. In this case, the impact can occur at a larger impact parameter and perihelion distance compared to the near side case, which will collide before perihelion. 

There is also a possibility that an apex-side impact occurs after perihelion.  In this scenario there are two post-perihelion impacts but once again the antapex impact is more likely to occur due to significantly larger impact parameter and perihelion distance.  In other words, impacts when the Earth is close to the antapex have  experienced the Solar gravitational focusing for longer; a similar case for bound objects was described in \citet{Moorhead2020}. They can therefore  arrive from a greater region in  space, increasing the number of them that can bend their trajectory and impact the Earth. In our simulations the mean perihelion distance for apex-side impacts is 0.51 au while for antapex-side impacts the mean perihelion is $q=0.77$ au. Additionally, the largest perihelion distance for apex-side impacts is 0.69 au while for antapex side impacts it is 1 au (which is the largest possible perihelion distance for an impact to occur anywhere).

\subsection{Latitudinal Distribution of Impacts}
In Figure \ref{fig:worldmap_flux} we show the relative flux of interstellar impactors on the surface of the Earth. To calculate this distribution we take into account the fact that a given radiant can impact at multiple locations on the Earth.  Specifically, a single radiant corresponds to an entire hemisphere of possible approach directions  comprised of orbits parallel to that radiant. To calculate this distribution we compute every latitude and longitude that correspond to the hemisphere defined by a given radiant, and then sum over all radiants. \added{The inclination of the Earth's pole with respect to the ecliptic is accounted for when  transforming between  ecliptic and equatorial reference frames. The gravitational influence of the Earth is not incorporated into this calculation. Figure \ref{fig:worldmap_flux}   is therefore only intended to depict a \textit{tentative} latitudinal dependence as opposed to an exact positional prediction.} Interstellar objects are more likely to hit the Earth at low latitudes close to the equator with a slight preference for the Northern hemisphere. \added{Figure \ref{fig:worldmap_flux} uses a Robinson projection.}

 \section{Discussion}\label{sec:conclusions}

In this paper we generated a synthetic population of Earth-impacting interstellar objects with the recently developed probabilistic method \citep{Marceta2023a}. The orders of magnitude increase in computational efficiency of this methodology enabled the generation of a statistically robust population of Earth-impactors.   In order to generate  ${\sim3.3\times10^4}$ Earth-impactors, we computed   ${\sim2.6\times10^{10}}$ synthetic overall objects.  

These simulations revealed several salient properties of the Earth-impacting interstellar objects which can be summarized as follows:

\begin{enumerate}
    \item Earth-impacting interstellar objects are most likely to approach from the solar apex and the galactic plane.
    \item The fastest Earth-impacting interstellar objects approach from the galactic plane and solar apex. 
    \item Earth-impacting interstellar objects  are typically  slower than the overall interstellar population.
    \item Earth-impacting interstellar objects typically have  hyperbolic, low-eccentricity orbits.
    \item  Earth-impacting interstellar objects are  likely to have orbits within the ecliptic plane, predominantly with retrograde motion. \added{Earth-impacting interstellar objects exhibit a sinusoidal inclination distribution for high $q$, similar to the overall population, but tend toward a uniform distribution for low $q$.}
    \item The fastest interstellar impactors collide in the spring when the Earth is moving towards the solar apex.
   
    \item Interstellar objects are more likely to impact the Earth in the winter when the Earth is in the direction of the antapex.
    
    \item Interstellar objects are  more likely to impact the Earth close to the equator and in the Northern hemisphere.

\end{enumerate}

Here we highlight caveats and areas of future works. These distributions are only applicable for interstellar objects that have  M-stars kinematics. Different assumed kinematics should change the distributions presented in this paper. The  salient features summarized in this section presumably also apply to different kinematics, perhaps to a muted or more distinct overall effect. It would be useful to perform similar analyses for different kinematics, although this is outside of the scope of this paper. For example, \citet{GreggWiegert2025} argued that interstellar meteors from $\alpha$-Centauri should be distinct in their average position and higher velocities  $\sim53$ km s$^{-1}$ at 1 au. Finally, these distributions are only applicable to objects sufficiently large to not be affected by gas drag in the interstellar medium.

In this paper we intentionally do not make any definitive predictions about the rates of interstellar impactors. In turn, we also do not make any claims regarding the presence or lack of interstellar meteors in extent data.  The distribution of Earth-impactors are open source, and  intended to be a useful tool for other researchers to investigate Earth-impacting interstellar objects of macroscopic sizes.

\textbf{Data Availability Statement:} The data affiliated with this publication is available at \url{https://doi.org/10.5281/zenodo.17693389}.

\section{acknowledgments}

\added{We thank the anonymous reviewer for insightful comments and constructive suggestions that strengthened the scientific content of this manuscript. We thank Mari\'a Hajdukov\'a for providing an informal review of the  manuscript upon submission. }

We thank Tom Statler, H\'ector Socas-Navarro, Brian O'Shea, Lia Coralles,  Adina Feinstein, Garrett Levine, Amanda Gill, and Jason Kohn for useful conversations. 

D.Z.S. is supported by an NSF Astronomy and Astrophysics Postdoctoral Fellowship under award AST-2303553. This research award is partially funded by a generous gift of Charles Simonyi to the NSF Division of Astronomical Sciences. The award is made in recognition of significant contributions to Rubin Observatory’s Legacy Survey of Space and Time.

D.M. acknowledges support by the Science Fund of the Republic of Serbia, GRANT No 7453, Demystifying enigmatic visitors of the near-Earth region (ENIGMA).

E.P.-A. acknowledges financial support from the LUMIO project, funded by the Agenzia Spaziale Italiana (2024-6-HH.0).

\section{contribution}
All authors contributed equally to this manuscript.
D.Z.S. led the paper writing and figure generation, and contributed to interpretation of data. D. M. performed the numerical simulations, led the interpretation of data, and contributed to figure generation. E. P. A. contributed to manuscript writing, interpretation of data, and figure generation.

\bibliography{sample7}{}
\bibliographystyle{aasjournal}

\end{document}